# Directivity patterns of laser-generated sound in solids: Effects of optical and thermal parameters

Victor V. Krylov

Department of Aeronautical and Automotive Engineering,
Loughborough University,
Loughborough, Leicestershire LE11 3TU, UK

## Abstract

In the present paper, directivity patterns of laser-generated sound in solids are investigated theoretically. Two main approaches to the calculation of directivity patterns of laser-generated sound are discussed for the most important case of thermo-optical regime of generation. The first approach, which is widely used in practice, is based on the simple modelling of the equivalent thermo-optical source as a mechanical dipole comprising two horizontal forces applied to the surface in opposite directions. The second approach is based on the rigorous theory that takes into account all acoustical, optical and thermal parameters of a solid material and all geometrical and physical parameters of a laser beam. Directivity patterns of laser-generated bulk longitudinal and shear elastic waves, as well as the amplitudes of generated Rayleigh surface waves, are calculated for different values of physical and geometrical parameters and compared with the directivity patterns calculated in case of dipole-source representation. It is demonstrated that the simple approach using a dipole-source representation of laser-generated sound is rather limited, especially for description of generated longitudinal acoustic waves. A practical criterion is established to define the conditions under which the dipole-source representation gives predictions with acceptable errors. It is shown that, for radiation in the normal direction to the surface, the amplitudes of longitudinal waves are especially sensitive to the values of thermal parameters and of the acoustic reflection coefficient from a free solid surface. A discussion is given on the possibility of using such a high sensitivity to the values of the reflection coefficient for investigation of surface properties of real solids.

## 1. Introduction

Laser generation of sound is a convenient non-contact method of sound generation in air, liquids, and solids. The possibility of generation of sound in solids by transient surface heating has been first described by White more than 50 years ago [1]. Since then, following rapid development of lasers as efficient non-contact sources of heat, this method of generating surface and bulk acoustic waves has been studied widely in respect of its applications to non-destructive testing of materials and to experimental solid state physics (see, e.g. monographs [2-4]). One should keep in mind that this method of elastic wave generation is truly non-destructive only in the so-called thermo-optical regime of generation that takes place if the intensity of laser light is not too high to cause the surface damage and material ablation.

To describe the behaviour of laser-generated sound in the thermo-optical regime of generation it is often assumed that, e.g. in the case of two-dimensional geometry of laser-illuminated area, the equivalent laser-induced source of acoustic waves (a line source) can be modelled as a dipole-type source comprising two horizontal forces applied to the surface in opposite directions (see Figure 1). The calculated radiation patterns of bulk longitudinal and shear acoustic waves generated by such a pair of forces are shown in Figure 2. In the case of three-dimensional geometry of a laser beam (a point source) a second pair of horizontal forces should be added in the perpendicular direction, which results in similar radiation patterns in vertical cross-sections.

This very simplified theoretical model has been introduced on the intuitive basis in the early 80-ies (see e.g. [2, 3, 5]), and it is still used by the theorists [6]. In the same time, a number of papers based on a rigorous approach to the developing of the theory of laser generation of sound in solids have been published in the eighties and nineties [7-13]. These papers, that take into account light absorption and thermal wave generation and propagation, show that there is more to the phenomenon than just a response of the medium to a pair of opposite forces. Attention to the effects of thermal and optical parameters of solids on laser generation of sound continued to be paid also in more recent publications (see e.g. [14, 15]).

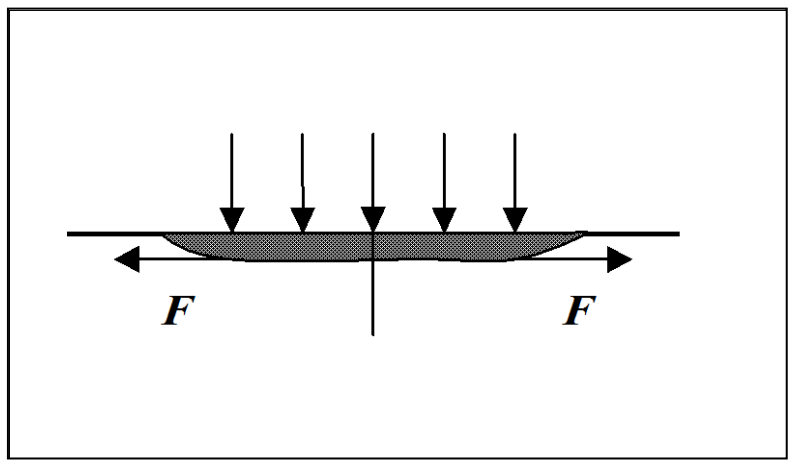

Figure 1. Modelling a laser-induced line source as a pair of horizontal forces.

In the present paper, the behaviour of directivity patterns of laser-generated sound in solids is examined on the basis of the rigorous approach developed in [7-12]. The theory takes into account all acoustical, optical and thermal parameters of a solid material and all geometrical and physical parameters of a laser beam. By analysing the expressions for laser-generated longitudinal, shear and Rayleigh waves, the applicability of the simplified dipole-source representation is discussed for all these cases. It is also shown that, for radiation in the normal direction to the surface, the amplitudes of longitudinal waves are very sensitive to the values of thermal parameters and of the acoustic reflection coefficient from a free solid surface. Finally, a discussion is given on the possibility of using such a high sensitivity to the values of the reflection coefficient for investigation of surface properties of real solids.

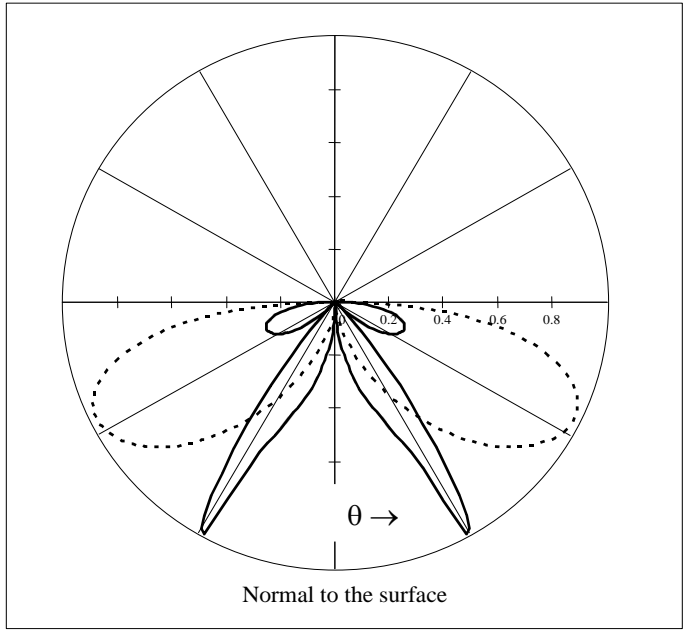

Figure 2. Dipole-type radiation patterns for a pair of horizontal forces applied to the surface: dashed curve corresponds to bulk longitudinal waves, and solid curve indicates shear waves.

## 2. Theoretical Background

Let us assume that a two-dimensional laser beam (a line source) is incident in normal direction onto the surface of a solid, which for simplicity is considered as elastically isotropic (see Figure 3). The problem of laser generation of sound in such a system can be described by the following governing equations [7-9].

The components of the displacement vector of laser-generated elastic waves $u_i$ must satisfy the equation of mechanical motion,

$$\rho \frac{\partial^2 u_i}{\partial t^2} = \sigma_{ij,j} \,, \tag{1}$$

and the linearised constitutive equation that takes into account temperature effects,

$$\sigma_{ij} = 2\mu u_{ij} + [\lambda u_{kk} - \gamma K(T-T_0)]\delta_{ij} \,. \tag{2}$$

Here $\sigma_{ij}$ are components of the elastic stress tensor, $\rho$ is the mass density of the medium, $u_{ij} = (1/2)(u_{i,j} + u_{j,i})$ are components of the linearised strain tensor, $\lambda$ and $\mu$ are the elastic Lame constants, $K = \lambda + 2\mu/3$ is the elastic compression modulus, $\gamma$ is the thermal expansion coefficient and $T_0$ is the initial temperature.

Equations (1) and (2) should be supplemented by the linearised equation of thermal balance, in which we ignore the effects of viscosity:

$$\rho c_v \frac{\partial T}{\partial t} - \kappa T_{,ii} + \gamma K T_0 \frac{\partial u_{ii}}{\partial t} = -\frac{\partial}{\partial z}[\beta I(x)f(t)exp(-az)]. \qquad (3)$$

Here $c_v$ is the material's specific heat at a constant volume, $\kappa$ is the coefficient of thermal conductivity, $\beta$ is the coefficient of laser light energy transmission into the medium, $\alpha$ is the light energy absorption coefficient, $I(x)$ is the spatial distribution of the intensity of laser radiation over the surface, and $f(t)$ describes the intensity modulation law. In what follows we assume for simplicity that modulation is time-harmonic, i.e. $f(t) = 1 + m\,cos\omega t$, and the modulation index $m$ is equal to unity.

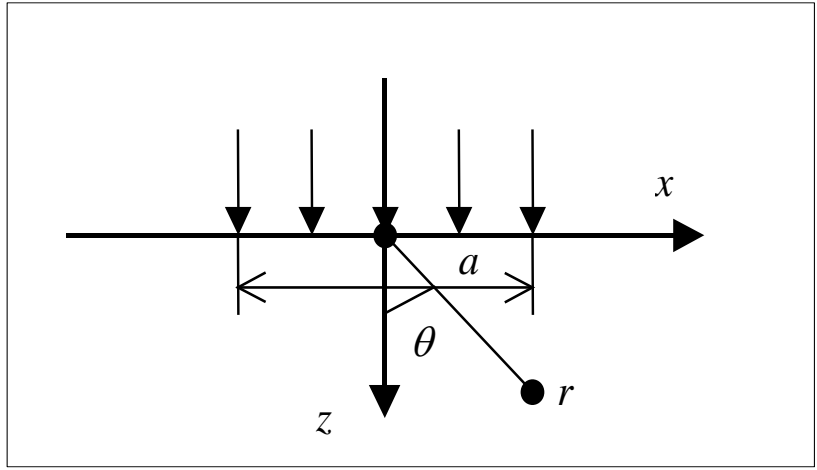

Figure 3. Geometry of the problem of laser generation of sound in solids.

The resulting field of generated acoustic waves must also satisfy the stress-free boundary conditions on the solid surface (at $z = 0$):

$$\sigma_{ij}\,n_j = 0, \qquad (4)$$

where $n_j$ are components of the normal unit vector to the surface. The generated temperature field must satisfy the condition of continuity of the thermal flow on the surface (at $z = 0$):

$$\frac{\partial T}{\partial z} = 0. \qquad (5)$$

Rigorous analysis of the system of simultaneous equations (1)-(3) and the boundary conditions (4), (5) is rather complex. A traditional simplification is in ignoring the small dilatational term $\gamma K T_0 \partial u_{ii}/\partial t$ in (3). In this case equation (3) does not contain mechanical displacements $u_i$ and can be solved independently versus $T$. Note that such an approximation is equivalent to ignoring the difference between isothermal and adiabatic values of the elastic constants $\lambda$ and $\mu$.

In what follows we limit our consideration to the case of laser generation of sound in metals and consider a typical case of the laser light penetration depth into the metal, $2\pi/\alpha$, being much smaller than the length of the thermal wave, $\lambda_T = 2\pi/k_T = 2\pi/(\omega\rho c_v/2\kappa)^{1/2}$. Then, the solution of (3) subject to the boundary condition (5) results in the following expression for $\partial T/\partial t$ away from the boundary:

$$\frac{\partial T}{\partial t} = [(1-i)k_T\beta\, I(x)/\rho c_v]\exp[-(1-i)k_T z - i\omega t] . \qquad (6)$$

Expressing particle vibration velocity in the medium $v_i = \partial u_i/\partial t$ via potentials $\varphi$ and $\psi$,

$$v_x = \frac{\partial \varphi}{\partial x} - \frac{\partial \psi}{\partial z} , \qquad (7)$$

$$v_z = \frac{\partial \varphi}{\partial z} + \frac{\partial \psi}{\partial x} , \qquad (8)$$

and substituting (7) and (8) into equations (1) and (2), using (6), one can obtain the following wave equations versus potentials (factor $exp(-i\omega t)$ is omitted):

$$\Delta\varphi + k_l^2 \varphi = [p/(\lambda + 2\mu)]I(x)exp[-(1-i)k_T z] = P(x,z), \qquad (9)$$

$$\Delta\psi + k_t^2 \psi = 0 . \qquad (10)$$

Here $p = (1-i)k_T\beta\gamma K/\rho c_v$, while $k_l = \omega/c_l$ and $k_t = \omega/c_t$ are wavenumbers of longitudinal and shear waves in the elastic medium, and $c_l = [(\lambda+2\mu)/\rho]^{1/2}$ and $c_t = (\mu/\rho)^{1/2}$ are their phase velocities.

Substituting (7) and (8) into the boundary conditions (4) and using equations (9) and (10), one can obtain the following homogeneous boundary conditions for the potentials:

$$k_t^2 \varphi + 2(\partial^2 \varphi/\partial x^2 - \partial^2 \psi/\partial x \partial z) = 0, \qquad (11)$$

$$k_t^2 \psi - 2(\partial^2 \varphi/\partial x \partial z - \partial^2 \psi/\partial z^2) = 0. \qquad (12)$$

This implies that the initial thermo-acoustical problem has been reduced to the solution of the equations (9) and (10) for potentials subject to the homogeneous boundary conditions (11), (12). It is obvious from (9)-(12) that the equivalent sources responsible for laser-generated sound are located in the bulk of the medium and not on its surface, which is described by the stress-free boundary conditions (11), (12).

The solution to the problem (9)-(12) can be obtained using the two-dimensional Green's function for an elastic half space [8]. The resulting formal expressions for $\varphi$ and $\psi$ include integration over wavenumber $k$ and over spatial coordinates $x$ and $z$:

$$\varphi(x,z) = -\frac{1}{4\pi} \int_{-\infty}^{\infty} \int_{-\infty}^{\infty} \int_{0}^{\infty} \left[ \frac{1}{v_l}(e^{-v_l|z-z'|} - e^{-v_l(z+z')})e^{ik(x-x')} - 8\frac{k^2 v_t}{F(k)} e^{-v_l(z+z')+ik(x-x')} \right] P(x',z')dkdxdz, \quad (13)$$

$$\psi(x,z) = -\frac{i}{\pi} \int_{-\infty}^{\infty} \int_{-\infty}^{\infty} \int_{0}^{\infty} \frac{k(2k^2 - k_t^2)}{F(k)} e^{-v_l z' - v_t z + ik(x-x')} P(x',z')dkdxdz. \quad (14)$$

Here $v_l = (k^2 - k_l^2)^{1/2}$, $v_t = (k^2 - k_t^2)^{1/2}$, and $F(k) = (2k^2 - k_t^2)^2 - 4k^2 v_l v_t$ is the so-called Rayleigh determinant. In the process of contour integration of (13) and (14) over the complex $k$-plain, the contribution of the poles gives the potentials of laser-generated Rayleigh surface waves propagating in both directions:

$$\varphi_R = -\frac{ip\Phi(k_R)}{F'(k_R)} \frac{4k_R^2 s}{(\lambda + 2\mu)[(1-i)k_T + q]} \exp(\pm ik_R x - qz) \quad (15)$$

$$\psi_R = -\frac{2k_R p\Phi(k_R)}{F'(k_R)} \frac{2k_R^2 - k_t^2}{(\lambda + 2\mu)[(1-i)k_T + q]} \exp(\pm ik_R x - sz) \quad (16)$$

Here $k_R$ is the wavenumber of Rayleigh waves that satisfies the equation $F(k_R) = 0$, the parameters $q$ and $s$ are defined as $q = (k_R^2 - k_l^2)^{1/2}$, $s = (k_R^2 - k_t^2)^{1/2}$, $F'(k_R)$ is the value of the derivative $dF(k)/dk$ of the Rayleigh determinant $F(k) = (2k^2 - k_t^2)^2 - 4k^2 v_l v_t$ taken at the point $k = k_R$, and $\Phi(k_R) = \int_{-\infty}^{\infty} I(x)\exp(-ik_R x)dx$ is the Fourier transform of the spatial distribution $I(x)$ of the laser light intensity over the surface.

The contributions of the saddle points in (13) and (14) give the potentials of laser-generated longitudinal and shear bulk waves propagating into the depth of the medium. At relatively large distances from the laser-illuminated surface the particle displacements in these waves are predominantly longitudinal, $u_r$, and shear, $u_\theta$, respectively. Using asymptotic relationships between particle displacements and potentials following from (7), (8) in a polar system of co-ordinates, $u_r = -(1/c_l)\varphi$, $u_\theta = (1/c_t)\psi$, the expressions for laser-generated longitudinal and shear bulk waves can be written in the form:

$$u_r(r,\theta) = -\frac{p\Phi(k_l \sin\theta)}{(\lambda + 2\mu)\sqrt{2\pi k_l r} c_l} \exp(ik_l r - i\pi/4) \left\{ \frac{k_l \cos\theta}{(1-i)^2 k_T^2 + k_l^2 \cos^2\theta} + \frac{i2k_l^4 \sin\theta \sin 2\theta[(k_t/k_l)^2 - \sin^2\theta]^{1/2}}{[(1-i)k_T - ik_l \cos\theta]F(k_l \sin\theta)} \right\}$$

(17)

$$u_\theta(r,\theta)=\frac{ip\Phi(k_t\sin\theta)}{(\lambda+2\mu)2\sqrt{2\pi k_t r}c_t}\exp(ik_t r-i\pi/4)\left\{\frac{k_t^4\sin 4\theta}{[(1-i)k_T+k_t(\sin^2\theta-k_l^2/k_t^2)^{1/2}]F(k_t\sin\theta)}\right\}$$

(18)

Directivity patterns of laser-generated longitudinal and shear waves are defined as $D_r(\theta) = |u_r(r,\theta)/u_{r,max}(r,\theta)|$ and $D_\theta(\theta) = |u_\theta(r,\theta)/u_{\theta,max}(r,\theta)|$ respectively, where $u_r(r,\theta)$ and $u_\theta(r,\theta)$ are determined by (17) and (18).

## 3. Numerical Calculations and Discussion

Let us analyse equations (17) and (18) for different relationships between the characteristic width of the laser beam $a$, the wavelengths of generated longitudinal and shear waves $\lambda_{l,t} = 2\pi/k_{l,t}$ and the wavelengths of thermal waves $\lambda_T = 2\pi/k_T$.

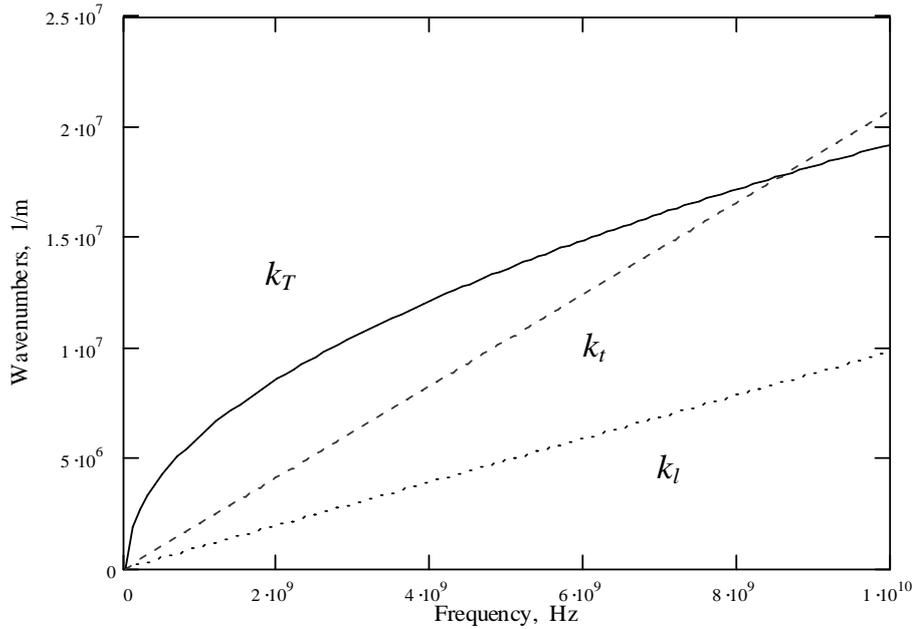

Figure 4. Wavenumbers of longitudinal, shear and thermal waves for Aluminium, $k_l$, $k_t$ and $k_T$ respectively, as functions of frequency.

For metals, at almost all frequencies of interest $\lambda_T \ll \lambda_{l,t}$, or $k_T \gg k_{l,t}$. Indeed, typical behaviour of $k_T$, $k_l$ and $k_t$ as functions of frequency $f = \omega/2\pi$ is illustrated in Figure 4 for Aluminium. One can see that $k_T \gg k_{l,t}$ for $f < 10^9$ Hz, i.e. for the whole range of frequencies used for non-destructive testing applications.

Using the above inequalities, let us first discuss the effect of thermal wave penetration into a solid on generation of shear bulk waves described by equation (18). Keeping in mind that $p = (1-i)k_T\beta\gamma K/\rho c_v$, one can see that in the main approximation the radiation pattern of shear waves does not depend on $k_T$, and for relatively narrow beams (when the effect of $\Phi(k_t \sin\theta)$ is negligible) it is defined largely by the factor $sin4\theta$ typical for dipole-type radiation from a pair of horizontal forces (see, e.g. [6]). Note that shear waves are not generated in normal direction simply because of the symmetry of the problem. Thus, for generated shear waves a thermo-optical source can be adequately represented as a dipole. The same is true also for generated Rayleigh surface waves described by equations (15) and (16).

However, for generated longitudinal waves (see equation (17)), the situation is essentially different. While the second term in figure brackets of (17) does not depend on $k_T$ for $k_T >> k_{l,t}$ and describes adequately the contribution of a dipole-type source, the first term does depend on $k_T$, and this limits significantly the applicability of dipole-source representation in respect of generated longitudinal waves. Obviously, the main influence of the first term occurs for longitudinal waves generated in the normal direction to the surface ($\theta = 0$) since the dipole-type contribution given by the second term vanishes at $\theta = 0$.

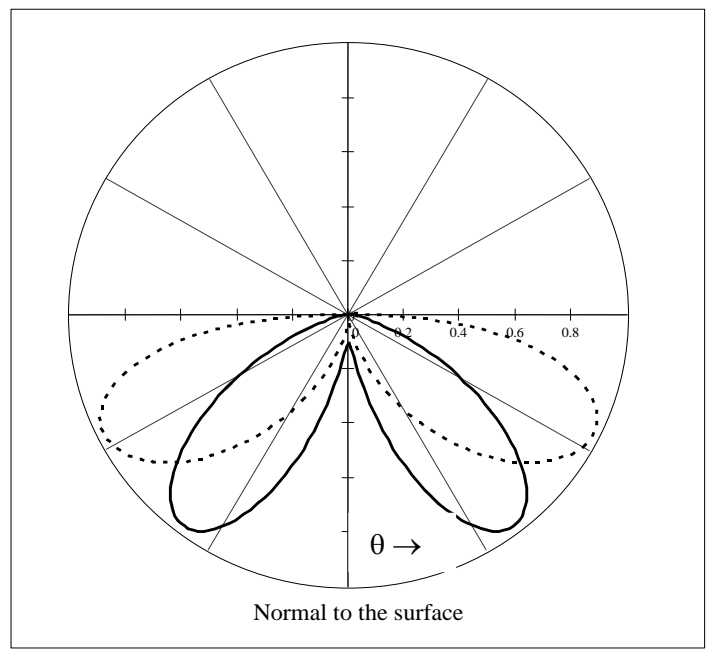

Figure 5. Directivity patterns of laser-generated longitudinal waves in Aluminium at frequency 5 MHz: $a = 0.1$ mm (dashed curve) and $a = 1.3$ mm (solid curve).

Typical radiation patterns of laser-generated longitudinal waves in Aluminium at frequency 5 MHz calculated according to equation (17) for two different values of laser beam width $a$ are shown on Figure 5: for $a = 0.1$ mm and $a = 1.3$ mm. As one can see, for very narrow laser beams ($k_{l,t}a << 1$; this corresponds to $a = 0.1$ mm in Figure 5) the relative amplitudes of longitudinal waves generated in normal direction ($\theta = 0$) are negligibly small as compared to the waves generated at oblique angles. The physical reason for this is that laser-generated sound field is a superposition of waves directly radiated into the bulk by a heated area and waves radiated towards the surface and reflected from it. In the normal direction the directly radiated and reflected waves nearly cancel each other out, the cancellation being almost complete for very small penetration lengths of thermal waves $l_T \approx \lambda_T = 2\pi/k_T$. At higher frequencies and for metals with higher values of $\kappa/\rho c_v$ the thermal wavelength $\lambda_T = 2\pi/k_T$ is larger, and radiation in the normal direction to the surface increases.

In the case of wider laser beams (this corresponds to $a = 1.3$ mm in Figure 5), the relative contribution of the first term in (17) responsible for radiation of longitudinal waves in the normal direction shows noticeable increase, whereas the second term in (17) becomes more suppressed by the factor $\Phi(k_l \sin\theta)$, in agreement with the so-called 'Product theorem' for large linear radiators and arrays of sources. For the same reason, the contribution of the whole expression (18) describing generation of shear waves is diminished due to the effect of the factor $\Phi(k_t \sin\theta)$. Note that the above-mentioned theoretical results are in good agreement with the existing experiments [10, 16].

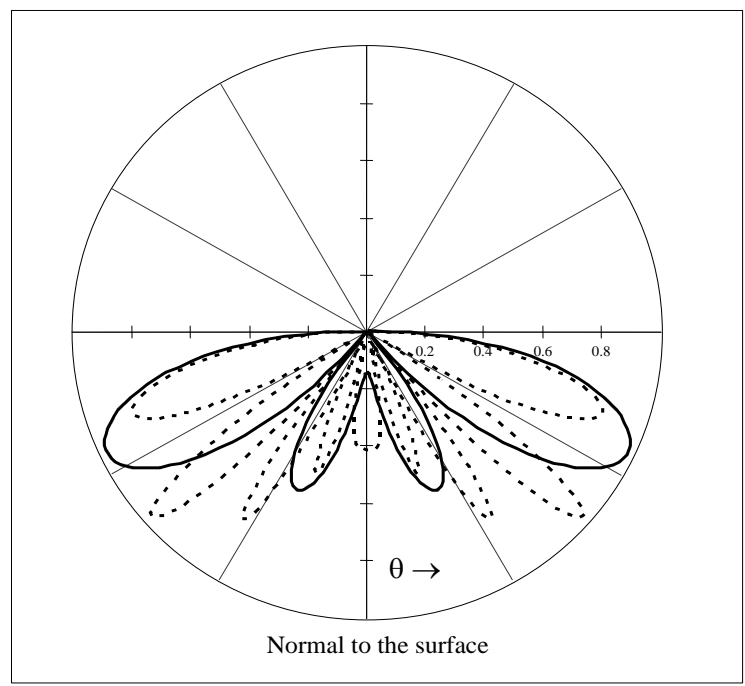

Figure 6. Directivity patterns of laser-generated longitudinal waves in Aluminium at frequency 5 MHz: $a = 2$ mm (solid curve) and $a = 6$ mm (dashed curve).

If to request that for a dipole-type radiation relative amplitudes of longitudinal waves in the normal direction should not exceed 10% of their maximum values at oblique angles, then for $a < \lambda_l = 2\pi/k_l$ the following practical criterion of dipole-source representation of laser-generated sound can be obtained from equation (17):

$$\frac{k_l}{k_T}\frac{k_l a}{\sin(0.43 k_l a)} < 0.08. \qquad (19)$$

If the condition (19) is not satisfied for given values of $k_l$, $k_T$ and $a$, then dipole-source representation is not adequate.

For wider laser beams, i.e. for $a > \lambda_l = 2\pi/k_l$, the contribution of longitudinal waves radiated in the normal direction is always significant due to the effect of the factor $\Phi(k_l \sin\theta)$ and can even be prevailing if $k_{l,t}a \gg 1$. Obviously, in this case no dipole-source representation is possible for laser-generated longitudinal waves. This situation is illustrated in Figure 6 showing calculated directivity patterns of laser-generated longitudinal waves in Aluminium at frequency 5 MHz for $a = 2$ mm and $a = 6$ mm. One can see that further increase of a laser beam width results in even more pronounced radiation of longitudinal waves in the normal direction and in general distortion of the initial directivity patterns due to the effect of the factor $\Phi(k_l \sin\theta)$. In the limit of an infinitely wide laser beam ($k_{l,t}a \to \infty$), which corresponds to a one-dimensional problem (only z-dependence remains), this reflects the well-known fact that under such circumstances only longitudinal waves can be generated by a laser source.

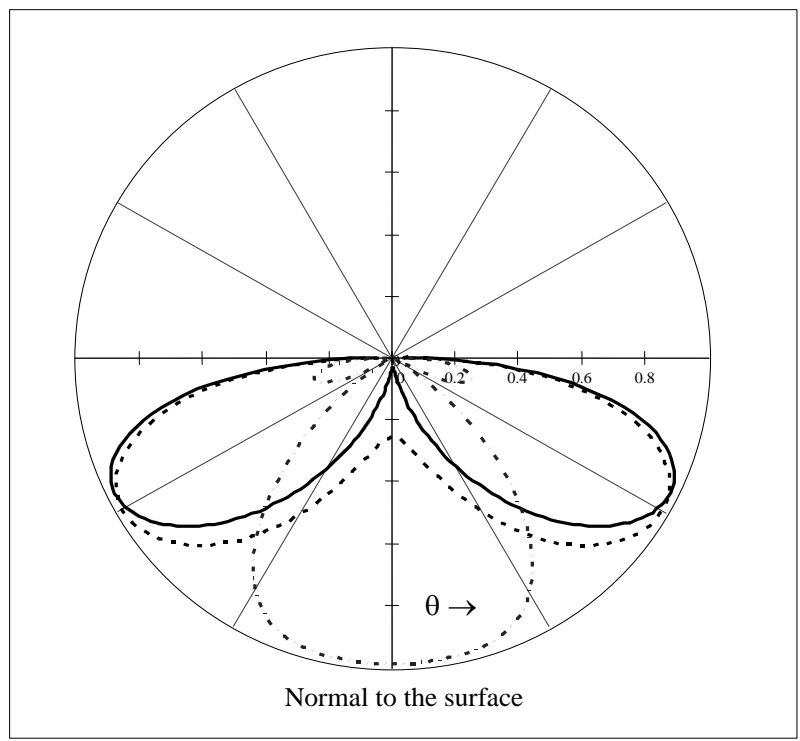

Normal to the surface

Figure 7. Directivity patterns of laser-generated longitudinal waves in Aluminium at frequencies 5 MHz (solid curve), 250 MHz (dashed curve) and 750 MHz (dash-dotted curve); $a = 0.01$ mm.

Figure 7 illustrates the effect of laser modulation frequency $f$ on directivity patterns of laser-generated longitudinal waves in Aluminium for a very narrow laser beam ($a = 0.01$ mm). Frequencies of generation are 5 MHz, 250 MHz, and 750 MHz. In agreement with the frequency dependence of $k_l$ and $k_T$ according to Figure 4, at frequencies 250 MHz and 750 MHz radiation of longitudinal waves in the normal direction becomes very significant, which means that dipole-source representation is invalid in such cases even for very narrow laser beams. This situation may be important in solid state physics applications of laser-generated sound [4], in particular for very short laser pulses having very wide frequency spectra (up to $10^9$–$10^{12}$ Hz).

## 4. Effect of Surface Properties of Solids on Laser-generated Sound

Referring to the results described in the previous sections, one should keep in mind that the possibility of dipole-source representation of laser-generated sound is associated not only with a relatively small penetration depth of thermal waves into the solid and a small width of a laser beam, but also with the fact that the reflection coefficient of elastic waves from a free solid surface in the direction normal to the surface is equal to minus one. This, of course, relates to the case of ideal free solid surfaces that were discussed above. Needless to say, however, that any modification of solid surfaces affecting their acoustic reflection coefficients (e.g., the presence of liquid droplets, thin films, surface roughness, etc.) can result in drastic changes in radiation patterns of laser-generated sound [9, 11, 12] that are incompatible with the dipole-source representation. The good thing is that such changes in radiation patterns of laser generated sound due to surface modifications can be used for investigation of physical and geometrical properties of real solid surfaces.

Indeed, according to [9, 12], the integral expressions for laser-generated acoustic waves in solids with ideal surfaces can be represented also in a slightly different form, which of course is equivalent to the above-mentioned equations (13) and (14). Namely, the expressions under the integrals in (13) and (14) can be rearranged to contain explicitly the reflection coefficients of longitudinal and shear acoustic waves, $V_{ll}(k)$ and $V_{lt}(k)$ respectively, from an ideal solid surface in the case of incidence of bulk longitudinal acoustic wave onto the surface at the angle $\theta$ defined by the expression $sin\theta = k/k_l$:

$$\varphi(x,z) = -\frac{1}{4\pi} \int_{-\infty}^{\infty} \int_{-\infty}^{\infty} \int_{0}^{\infty} \left[ \begin{array}{c} e^{-\nu_l |z-z'|+ik(x-x')} + \\ V_{ll}(k) e^{-\nu_l(z+z')+ik(x-x')} \end{array} \right] \frac{1}{\nu_l} P(x',z') dk dx dz, \qquad (20)$$

$$\psi(x,z) = -\frac{1}{4\pi} \int_{-\infty}^{\infty} \int_{-\infty}^{\infty} \int_{0}^{\infty} V_{lt}(k) e^{-\nu_l z' - \nu_t z + ik(x-x')} \frac{1}{\nu_l} P(x',z') dk dx dz. \qquad (21)$$

Here $V_{ll}(k)$ and $V_{lt}(k)$ are the corresponding reflection coefficients for an ideal surface defined by the expressions

$$V_{ll}(k) = -\frac{(2k^2 - k_t^2)^2 + 4k^2 \nu_l \nu_t}{F(k)}, \qquad (22)$$

$$V_{lt}(k) = -\frac{4ik\nu_l(2k^2 - k_t^2)}{F(k)}, \qquad (23)$$

where $F(k) = (2k^2 - k_t^2)^2 - 4k^2 \nu_l \nu_t$ is the earlier mentioned Rayleigh determinant.

If the surface is modified, e.g. by the presence of thin films, the reflection coefficients (22) and (23) must be replaced with the new reflection coefficients $\underline{V}_{ll}(k)$ and $\underline{V}_{lt}(k)$ taking into account the effects of modified surface properties. These new reflection coefficients should be substituted into equations (20) and (21) to be used for further analysis.

Some preliminary investigations in this direction have been reported in [9, 12] for surface modifications associated with the effects of deposited thin films or damaged subsurface layers. If deposited films or damaged subsurface layers are much thinner than wavelengths of laser-generated acoustic waves, their effects on the reflection coefficients can be characterised by such integral parameters as surface elasticity, surface mass density and surface tension [12]. As expected, the main influence of these surface parameters on laser generation of sound occurs in the case of longitudinal waves radiated in the normal direction, which is very sensitive to surface modifications.

In the paper [11], initial estimates of the effects of surface roughness on laser-generated sound in solids have been carried out. For statistically rough surfaces, the parameter of interest is the averaged coherent field of laser-generated acoustic waves. For its description, the above-mentioned reflection coefficients for an ideal flat surface described by equations (22) and (23), $V_{ll}(k)$ and $V_{lt}(k)$, should be replaced with the corresponding reflection coefficients $<V_{ll}(k)>$ and $<V_{lt}(k)>$ for statistically averaged fields that can be obtained as a result of the solution of the statistically defined problem. A general solution to such a problem for elastic media with statistically rough surfaces is very cumbersome. Therefore, only the simplest case of laser-excited waves radiated in the normal direction to the mean (flat) surface (at $\theta = 0$) was considered in [11]. For this particular case, one can use the analogy with a scalar problem of sound reflection from statistically rough surfaces of liquids [17], which allows one to obtain only the 'longitudinal' reflection coefficient in the normal direction, $<V_{ll}(0)>$. Fortunately, as was mentioned above, the case of longitudinal waves radiated in the normal direction is the most interesting for practical applications. If a statistically rough surface is characterised by the mean-square height $\sigma$ and by the correlation length $a$, the corresponding reflection coefficient $<V_{ll}(0)>$ describes the effect of these surface roughness parameters on laser-generated longitudinal acoustic waves radiated in the normal direction. The initial calculations show [11] that the effect of surface roughness may result in noticeable amplification of longitudinal waves generated in the normal direction, which can be used for non-contact evaluation of the parameters of surface roughness.

## 5. Conclusions

In the present paper, the directivity patterns of laser-generated sound have been examined on the basis of the rigorous theory that takes into account all acoustical, optical and thermal parameters of a solid material and all geometrical and physical parameters of a laser beam. It has been demonstrated that the widely used dipole-source representation of laser-radiated sound is rather limited, especially in respect of generated longitudinal waves. A practical criterion has been established to define the conditions under which the dipole-source representation gives predictions with acceptable errors.

Although the dipole-source representation for laser-generated sound may be helpful for understanding directivity patterns of radiated acoustic waves in some simple practical situations, it cannot describe many important cases of radiation of longitudinal waves in the normal direction to the surface. The rigorous approach accounting for optical and thermal parameters of solids described in this paper encompasses all possible relationships between geometrical and physical parameters of a solid and of a laser beam. The obtained results for directivity patterns of laser-generated longitudinal acoustic waves and their strong dependence on surface properties of solids can be used for non-contact investigation of real solid surfaces.